\documentclass[%
 reprint,
superscriptaddress,
 amsmath,amssymb,
 aps,prf
]{revtex4-2}

\usepackage{graphicx}
\usepackage{newtxtext}
\usepackage{newtxmath}
\usepackage{natbib}
\usepackage{hyperref}
\hypersetup{
    colorlinks = true,
    urlcolor   = blue,
    citecolor  = black,
}

\newcommand{\RomanNumeralCaps}[1]
\linenumbers
\DeclareMathOperator\Ca{Ca}
\DeclareMathOperator\erf{erf}

\begin{document}
\title{Freezing receding contact lines}

\author{Rodolphe Grivet}
 \affiliation{Laboratoire d'Hydrodynamique (LadHyX), UMR 7646 CNRS-Ecole Polytechnique, IP Paris, 91128 Palaiseau CEDEX, France}%
\author{Axel Huerre}
 \affiliation{MSC, Universit\'e de Paris, CNRS (UMR 7057), 75013 Paris, France}%
 \author{ Thomas Séon}
\affiliation{Institut Jean Le Rond $\partial$’Alembert, UMR 7190, CNRS-Sorbonne Université, 75005 Paris, France}
\author{Laurent Duchemin}
\affiliation{Laboratoire de Physique et Mécanique des Milieux Hétérogènes (PMMH), CNRS- ESPCI Paris- Université PSL-Sorbonne Université-Université Paris CIté, Paris, France}
 \author{Christophe Josserand}
 \email{christophe.josserand@ladhyx.polytechnique.fr}
  \affiliation{Laboratoire d'Hydrodynamique (LadHyX), UMR 7646 CNRS-Ecole Polytechnique, IP Paris, 91128 Palaiseau CEDEX, France}%

\begin{abstract}
We investigate experimentally the receding of a contact line when a Peltier module is pulled out of a water bath at constant speed, while a ice layer is also growing at constant speed on the Peltier module.
A steady regime is therefore reached for all the parameters used in this studied, corresponding to a dynamical stationnary meniscus. We show that the height of the meniscus provides most of the properties of the
flow. For high pulling rate, it is related to the amount of liquid of the equivalent Landau-Levich (LL) film that would be extracted from the bath, which is eventually freezing as the plate is lifted upward. For smaller velocity, so that
no LL film would be formed without freezing, the meniscus height is directly linked to the contact angle of water on ice in these conditions. Solving numerically the meniscus equation taking into account the solidifcation of water,  
our results suggest that the contact angle of water on ice should be around $6^\circ$.
\end{abstract}

\maketitle

\section{Introduction}
\label{sec:intro}

The interaction between wetting phenomena and solidification has been the object of several recent experimental \citep{Schiaffino1997,Tavakoli2014,DeRuiter2017, Koldeweij2021, Grivet2022} and theoretical \citep{Herbaut2020} studies aiming to better understand the coupling between both physical processes. This coupling indeed gives rise to unexpected features and patterns in various situations, such as the stick-slip motion of a contact line when a liquid flow is forced on a cold surface \citep{Herbaut2019}, or the spontaneous retraction of water on ice after the impact of a water drop on a solid below 0 \textcelsius \citep{Thievenaz2020b}. These features of solidifying capillary flows are of interest in various area, from aircraft icing \citep{Cebeci2003} to the understanding of environmental flows \citep{Demmenie2023}. Despite this interest and these studies, few is in fact known on the particular nature of the interaction between the viscous-capillary flow near the contact line and the solidification.

In fact, the simpler single question of contact line motion has been the motivation of numerous fundamental studies at least since \citet{Huh1971}, who exhibited that the classical hydrodynamic description of fluids becomes invalid at small scales near the moving contact line due to the divergence of the viscous dissipation. This observation stressed out the need for small scales models involving some sort of cut-off length \citep{Voinov1976,DeGennes1986,Bonn2009, Snoeijer2013}, among which one can cite the use of a Navier slip length or the consideration of Van-der-Waals interaction \cite{Eggers2004}. The use of a cut-off length has been shown to capture well the interface deformations associated with the motion of the contact line \citep{Marsh1993}. More recently, these models were also used to understand some macroscopic features of dynamic wetting situations, such as the departure to Landau-Levich films in the case of receding contact lines \citep{Eggers2005, Snoeijer2006b}. Quite generally, its is usually assumed that the liquid-gas interface meets the solid surface with a given angle, a property of the materials considered, usually considered to be the Young-Dupré equilibrium angle. At larger scales, this interface is deformed by the viscous forces, and then connects to a macroscopic profile dictated by the broader equilibrium at stake, for instance between gravity and capillarity for the case of dynamic menisci.

For the particular case of water moving on ice, using this framework implies assuming the existence of a contact angle between water and ice, a fact that is not trivial, especially in situations when solidification is involved, and hence when the system is out of thermal equilibrium. The wettability of ice-water systems has itself been the topic of few studies since \citet{Knight1967}, who proposed experimental evidence that water would not completely wet ice. Since then, the quest for obtaining a contact angle between water and ice has mainly been focused on trying to experimentally study near 0 \textcelsius \: situations in which no heat transfer of phase change is involved, hence allowing to measure that contact angle using classical tools similar to sessile drop analysis \citep{Knight1971, Ketcham1969, Makkonen1997, Thievenaz2020b, Demmenie2023}. An other study \citep{Makkonen1997} aimed at measuring independently the different surface energies involved in the Young-Dupré equation in order to deduce the value of the equilibrium contact angle. These different studies provide values for the contact angle between water and ice ranging from "zero or very near zero" \citep{Knight1971} to 40$^\circ$ \citep{Makkonen1997} with no strong consensus in between. These discrepancies might be due essentially to the anisotropic nature of ice \citep{Libbrecht2017}, or to the difficulty of removing experimentally all heat transfer. 

Despite this strong divergence of the various measurements made near equilibrium, the experimental observation of spontaneous dewetting of water on growing ice firstly noted by \citet{Knight1967} and repeated by \citet{Thievenaz2020b} in a different experimental situation still suggests that water only partially wets ice and still remains unexplained. In this article, we hence wish to force a well controlled thermally out of equilibrium situation and observe the receding of water on ice. Due to the unsteady nature of the retraction of a drop on a solid, we move to the configuration of the dynamic menisci, in which a vertical plate is withdrawn from a liquid bath (surface tension $\gamma$, dynamic viscosity $\mu$ and density $\rho_l$) at a constant velocity $U_p$. This situation is expected to be stationary when no phase change is involved and when the capillary number of the system is sufficiently low \citep{Snoeijer2006b}.

More precisely, \citet{Eggers2004a,Eggers2005} showed that a stationary contact line can exist as long as the plate capillary number does not exceed a critical value, $\frac{\mu U_p}{\gamma}=\Ca_p<\Ca_{cr}(\theta_e,\lambda)\propto \theta_e^3$ where $\theta_e$ is the equilibrium contact angle imposed at the intersection of the plate and the free surface and $\lambda$ is the Navier slip length. This is due to the fact that the outer static meniscus, described by the Laplace gravito-capillary equation, has a maximum possible curvature reached when the moving contact line is located at a height $Z_{cl}=\sqrt 2 l_\gamma$ over the bath, $l_\gamma=\sqrt{\frac{\gamma}{\rho_l g}}$ being the capillary length. On the other side, \citeauthor{Eggers2005} shows that the inner viscous-capillary solution imposes a given curvature far from the contact line that increases with the capillary number. Therefore, when $\Ca_{cr}$ is exceeded, both regions can not be matched. In that case, \citet{Snoeijer2006b} showed that the contact line is entrained by the plate while slowly slipping on it, giving rise to the coexistence of two flat liquid films above the static meniscus, the lower one being exactly the one studied by \citet{Landau1942}. As time goes, this film gradually occupies all the plate above the static meniscus, and its thickness is expected to be $H_{LL}=\alpha_{LL} l_\gamma  \Ca_p^{2/3}$, where $\alpha_{LL}=0.94...$ is a numerical constant arising from the matching procedure.

In this article, we study the configuration in which the plate is made of a layer of ice and a constant solidification velocity is imposed in the direction perpendicular to this moving ice plate. Our experiments allow to vary in a controlled way the solidification and the plate upwards velocities, while measuring the height of the contact line relative to the bath. We find that this situation is stationary at any value of the capillary number over almost two decades and propose a simple scaling law based on the physics of isothermal Landau-Levich films to explain all experiments in which the contact line was observed above its maximum static position $\sqrt{2}l_\gamma$. To understand the rest of our observations, we then propose a modification of the classical thin film equation that allows to account for solidification, and whose integration relies on imposing a value for the ice-water contact angle at the contact line. Using numerical methods, we integrate this model in order to find an effective value of the contact angle between water and ice that allows to reproduce all of our data. 

The paper is organized as follows. In \S \ref{sec:exp}, we present the experimental setup and results. A particular care is given to the un-obvious measurement of the solidification front velocity. In \S \ref{sec:model}, we derive the model for the thin film equation with solidification, and present the numerical methods and results, before drawing some concluding discussions.

\section{Experiments}
\label{sec:exp}

\subsection{Experimental set-up}

\begin{figure}
  \centerline{\includegraphics[width=0.55\textwidth]{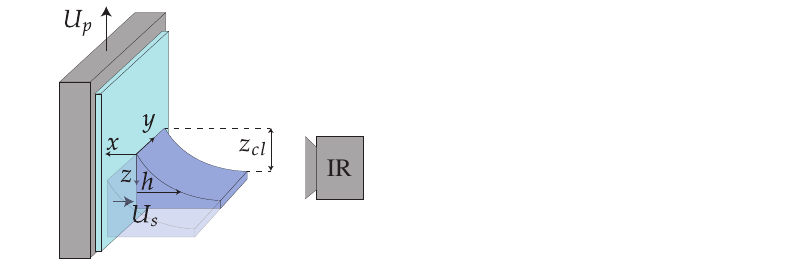}}
  \caption{Sketch of the experimental set-up. The non-dimensional coordinates $z$ is oriented downwards, $z=0$ denoting the contact line and $z=z_{cl}$ the bath.}
\label{fig:setup}
\end{figure}

A sketch of the experimental set-up is shown on Figure \ref{fig:setup}. It consists of a vertical 4 x 4 cm Peltier module connected to a tension generator and set to a translation stage. The hot side of the Peltier module is cooled by a cryostat, while the cold side is exposed to the bath. The bath is made of pure degassed water, maintained at $T_l \approx 5 $ \textcelsius\: by placing it in a mixture of brine and solid ice. During an experiment, the Peltier module is immersed in the bath, a constant tension $E$ is then fed to the circuit so that the module absorbs a constant (unknown) thermal power $\phi_p$ (units $\rm{W\cdot m^{-2}}$) at its cold side exposed to liquid water. Immediately after, a layer of ice appears and the module is lifted upwards at a constant velocity $U_p\in [0.5\:;\:8]\:\rm{mm \cdot s^{-1}}$. The motion is observed using an infrared camera at variable framerates depending on the plate velocity and with a spatial resolution of $45 \: \rm{\mu m \cdot px^{-1}}$, hence around 1\% of the capillary length. The level of the bath is manually measured for each experiment: when the plate is lifted upwards, it is defined as the position of the pinned contact line when it is first detectable on the infrared images (as when it is below the level of the bath, it is hidden to the camera by the downwards meniscus). This measurement allows a precision of $\pm 4 \: \rm{px}$.

\subsection{Contact line detection}

\begin{figure}
  \centerline{\includegraphics[width=0.45\textwidth]{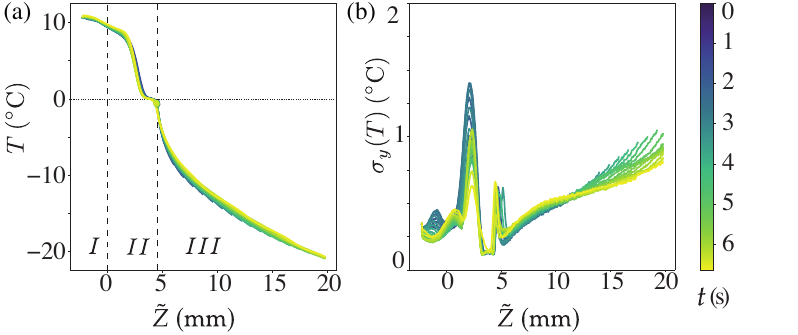}}
  \caption{(a) Temperature profile along the vertical axis at different timesteps, averaged along the $y$-axis. (b) Standard deviation of the temperature along the $y$-axis.}
\label{fig:profils_temp}
\end{figure}

In what follows, lengths indicated by capital letters will be dimensional, and lower case ones will be their non-dimensional counterparts using the capillary length $l_\gamma=2.77 \:\rm{mm}$ at $0$~\textcelsius \: for water. On Figure \ref{fig:profils_temp} (a), the average temperature along the $y$ axis is shown for a given experiment ($U_p=4 \:\rm{mm \cdot s^{-1}}$) as a function of the height relative to the bath, $\tilde{Z}=Z_{cl}-Z$, while Figure \ref{fig:profils_temp} (b) shows its standard deviation along $y$. The curves are colour-coded by the time. Firstly, the good superposition of all curves show that a stationary regime is quickly obtained. Secondly, the low magnitude of the variations of the temperature along the $y$-axis alows to consider only the two-dimensional problem. Finally, we temperature profile along $\tilde{Z}$ can be separated in three different regions: temperatures measured in region (I), below the level of the bath, are only reflections from the surrounding ambient temperature room; above the level of the bath, the temperature decreases down to $0$ \textcelsius , and then undergo a brutal change in slope which indicates the transition between the wetted region (II) and the ice in region (III). As observed on the profiles, the temperature continues to decrease in region (III) due to the continuous cooling imposed by the Peltier module.

We observe that the contact line position is in fact stationary throughout all experiments, a consequence of the good superposition of the curves of Figure \ref{fig:profils_temp}, which allows to characterize each experiment by a single value for $z_{cl}$.

\subsection{Solidification rate measurement}

The constant tension imposed to the Peltier module implies that the module will absorb a constant thermal power per unit surface $\phi_p$ that allows to grow a layer of ice of thickness $e(t)$ in the $x$ direction. The heat removal serves either to cool down the existing ice or to remove latent heat and increase $e$. Therefore, a constant solidification rate is expected only when the former is negligible compared to the latter. Both energies can be evaluated assuming a linear temperature profile in the ice. Under this assumption, the sensible heat and the latent heat removed to grow a layer of thickness $e$ write:

\begin{equation}
E_s\sim \rho_i c_{p,i} e \frac{\phi_p}{k_i}e\sim \frac{\phi_p}{D_i}e^2
\text{    and    }
E_l\sim \rho_iL_fe
\end{equation}

Where $\rho_i$, $c_{p,i}$, $k_i$, $D_i$  are the ice density, massic heat capacity, thermal conductivity and diffusivity, and $L_f$ is the latent heat of fusion of water. 

In fact, we can expect a linear growth of the ice layer when most of the heat removal is used as latent heat removal, hence when $E_s<<E_l$, which can be rewritten as:
\begin{equation} \label{eq:us}
e=U_s t \text{  where   } U_s=\frac{\phi_p}{\rho_i L_f}   \text{    and    } t<<\tau_s=\frac{D_i}{U_s^2}
\end{equation} 

In this article, the investigated solidification rates will be at most $U_s\approx 10^{-4} \rm{\:m\cdot s^{-1}}$, hence $\tau_s\approx 100\rm{\:s}$, so that a constant solidification rate hypothesis is always justified. This scaling approach can be formalised by solving the full Stefan problem with constant heat flux imposed at the bottom of the ice and expanding the solution in powers of $t/\tau_s$, giving a similar conclusion on the characteristic time of linear growth (see for instance \citet{Tao1979}).

In fact, the solidification rate can be measured directly for each experiment using the decrease of the temperature at the surface of the ice in region (III) of Figure \ref{fig:profils_temp} (a). Assuming a linear temperature profile in the $x$ direction within the ice with a slope  $\sigma$, in region II where the water film is still there, the front velocity is simply $U_s=-\frac{k_i \sigma}{\rho_i L_f}$. Following a material point at the surface of the ice, the temperature at this point is the ice melting temperature $T_f$ while it stays below the contact line position, and when above, it is exposed to air and hence the heat flux at this point vanishes causing the strong decrease of the surface temperature. This situation can be modelled at short times in the reference frame moving upwards at velocity $U_p$ with $t=0$ the time at which the material point is at the contact line. Neglecting the vertical and lateral heat difusions towards the horizontal one in the thin ice layer, the temperature difference $\bar{T}=T_f-T$ exhibits a self-similar solution of the form $\bar{T}=\sigma x f\left(\frac{x^2}{D_i t}\right)$ where $f$ is an unknown function. Solving the diffusion equation for $f$ using appropriate boundary conditions  ($\partial_x\bar T(x=0)=0$ and $\partial_x\bar T(x\rightarrow +\infty)=\sigma $) gives the temperature profile within the ice:

\begin{equation}
\bar{T}(x,t)=\sigma \Big [\sqrt{\frac{4D_it}{\pi}}e^{-x^2/(4D_it)}+ x \erf\Big(\frac{x}{2\sqrt{D_i t}}\Big) \Big ] 
\end{equation}

Which allows to simply express the ice surface temperature as:

\begin{equation}
T(0,t)=T_f-\sigma \sqrt{\frac{4D_it}{\pi}}
\end{equation}

By fitting this prediction to the temperature profiles in region (III) of Figure \ref{fig:profils_temp} (a), the value of the temperature gradient $\sigma$ is obtained at each timestep, and hence the solidification rate is measured. 
The solidification rates are measured to be constant for each experiments, a consequence of the good superposition of the curves of Figure \ref{fig:profils_temp}, which allows to characterize each experiment by a single value for $U_s$.

\subsection{Experimental results}

\begin{figure}
  \centerline{\includegraphics[width=0.45\textwidth]{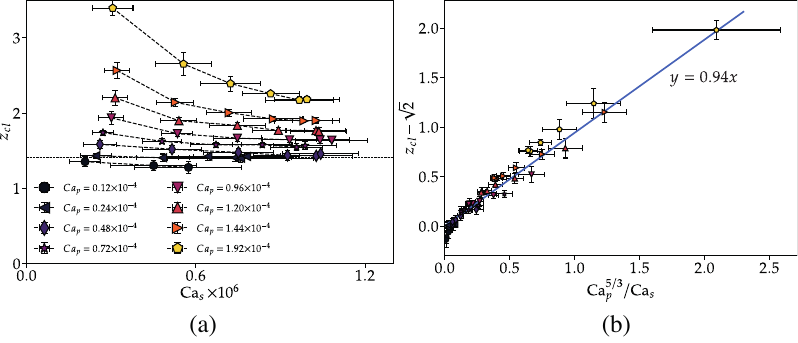}}
  \caption{(a) Contact line position as a function of the solidification capillary number for different plate capillary numbers. The dashed horizontal line represents $z_{cl,c}=\sqrt{2}$. (b) Contact line over-elevation as a function of the composite non-dimensional number $\Ca_p^{5/3}/\Ca_s$. The straight line is the one predicted using the coefficient from \citet{Landau1942}.}
\label{fig:posca}
\end{figure}

Each experiment is now characterised by the imposed parameters $(U_s,U_p)$ and the measured $z_{cl}$. The two velocities are non-dimensionalised by the viscous-capillary velocity $\gamma/\mu$ as: $Ca_p=\frac{\mu U_p}{\gamma}$ and $Ca_s=\frac{\rho_i}{\rho_l}\frac{\mu U_s}{\gamma}$. On Figure \ref{fig:posca} (a), the non-dimensional position of the contact line is shown as a function of the solidification capillary number for all plate capillary numbers. Clearly, the contact line rises as the plate capillary number is increased, while it decreases when solidification rate increases.

Most remarkably, it can be noticed that in the present experimental set-up, contact lines can recede at a constant velocity at heights higher than the isothermal critical value of $z_{cl,c}=\sqrt{2}$. In the isothermal case, these contact lines should rise to infinity leaving a Landau-Levich film \citep{Landau1942} behind. However in the present situation, due to solidification, the liquid in the film is gradually solidified so that eventually the free surface meets the ice. This can be modelled by noticing that the mass flux fed to the LL film at the top of the static meniscus is known to be $J_{LL}=\alpha_{LL} \rho_l l_\gamma U_p \Ca_p^{2/3}$. Applying a mass balance on the residual height above the static meniscus leads to:

\begin{equation}
\label{eq:scaling1}
\alpha_{LL} \rho_l l_\gamma U_p \Ca_p^{2/3}=(Z_{cl}-\sqrt{2}l_\gamma)\rho_i U_s,
\end{equation}

so that by rearranging the terms we obtain:

\begin{equation}
\label{eq:scaling2}
z_{cl}-\sqrt{2}=\alpha_{LL} \frac{\Ca_p^{5/3}}{\Ca_s}
\end{equation}

Note that a similar idea was proposed earlier in the case of dewetting of porous media by \citet{Raphael1999} and \citet{Aradian2000}, but without experimental evidence. Here, the prediction \ref{eq:scaling2} is tested against the experimental data on Figure \ref{fig:posca} (b) without any free parameter. It is clear that the experiments follow the behaviour predicted by this simple approach.

The predictive law \ref{eq:scaling2} is obviously only valid for experiments that would be in the entrained film regime in isothermal conditions, and hence does not explain our experiments where $z_{cl}<\sqrt{2}$. We now move to a more complete model that contains all the physics at stake.

\section{Modelling}
\label{sec:model}

\subsection{Hydrodynamic modelling}

The hydrodynamic model is developed under the assumption of small slopes $h'(z)<<1$. This is usually true in the viscous capillary region but obviously wrong in the static meniscus. Nonetheless, as pointed out by \citet{Eggers2005}, the small slope hypothesis is no longer needed for this region as the shape is dictated by an equilibrium between gravity and capillarity alone. Subsequently, the usual thin film equations can be used, which allows to compute the volumetric flux along the plate and oriented downwards as:

\begin{equation}\label{eq:flux}
J=\Big (\frac{\gamma}{\mu} K' +\frac{\rho_l g }{\mu}\Big) (H^3/3+\Lambda H^2)-U_pH
\end{equation} 

where $K$ is the interface curvature and $\kappa$ its non-dimensional counterpart. Following \citet{Eggers2005} and \citet{Snoeijer2006b}, we introduced the slip length $\Lambda$ to avoid the divergence of the viscous dissipation at the contact line. For systems close to perfect wetting, it is expected that this slip length is of the order of the molecular size \citep{Barrat1999}.

Now, it is possible to formally introduce the effect of solidification using a local mass balance in a similar way as previously, by noting that:

\begin{equation}\label{eq:solid}
\rho_l\partial_Z J=-\rho_i U_s
\end{equation}

which effectively models solidification as a mass leakage and hence is equivalent to the modelling of a thin film flowing on a porous medium with constant pumping \citep{Aradian2000}.

Using the final boundary condition that $J(Z=0)=0$, meaning that the contact line is not advancing or receding in the fixed reference frame of the laboratory, equation \ref{eq:solid} can be integrated as:

\begin{equation}
J=-\frac{\rho_i}{\rho_s} U_s Z
\end{equation}

Thus, using Equation \ref{eq:flux} and introducing the ratio $\phi=\frac{Ca_s}{Ca_p}$, we obtain in the dimensionless variables the following equation for the solidifying dynamical meniscus:

\begin{equation}\label{eq:lub}
\kappa'+1=3Ca_p\frac{1-\dfrac{z}{h}\phi}{h^2+3\lambda h}.
\end{equation}

This third order differential equation should be integrated with three boundary conditions obtained at the bath level and at the contact line:

\begin{equation} \label{eq:CAL}
\left\{
\begin{array}{c}
  h(z=0)=0\\
  h'(z=0)=\theta_e\\
  h'(z\rightarrow+\infty)=+\infty.\\
\end{array}
\right.
\end{equation}
It is worth noting that in our experiments, we fix $Ca_p$ and $Ca_s$, and we measure $z_{cl}$ so that resolving this set of equation will allow to deduce $\theta_e$ that will be therefore our free parameter.

\subsection{Numerical integration}

The system \ref{eq:lub}-\ref{eq:CAL} should be integrated numerically to obtain predictions on the contact line height depending on the free parameter $\theta_e$ and achieve comparison with the experiments. This requires some analytical work as $\kappa'$ goes to infinity in $z=0$ due to the boundary conditions. 

In this, we follow \citet{Eggers2005} to obtain an asymptotic solution of equation \ref{eq:lub} close to the contact line on scales of order $\lambda$. 
To put it in a nutshell, the lengths $h$ and $z$ are rescaled as $\hat{h}=h/(3\lambda)$ and $\hat{z}=\theta_e z/(3\lambda)$. Then, the small parameter  $\epsilon=3\Ca_p/\theta_e^3$ is introduced, and the interface profile is expanded as $\hat{h}=\hat{z}+\epsilon \hat{h}_1$, $ \hat{h}_1$ being of order 1. Plugging this expansion in \ref{eq:lub} and solving for the different orders in $\epsilon$ finally yields the solution close to the contact line, $\hat{h}_\lambda$:

\begin{equation}\label{eq:inner}
\begin{array}{c}
\hat{h}_\lambda =  \hat{z}+\epsilon (1-\epsilon^*)\Big(\frac{\hat{z}^2}{2}\ln (\hat{z})-\frac{\hat{z}^2}{4}-\frac{(\hat{z}+1)^2}{2}\ln (\hat{z}+1)\\
+\frac{(\hat{z}+1)^2}{4}+C\frac{\hat{z}^2}{2}-\frac{1}{4} \Big)
\end{array}
\end{equation}

where $\epsilon^*=\phi/\theta_e$ and $C$ is an unknown constant. This solution verifies the two first boundary conditions \ref{eq:CAL}, while containing a free parameter $C$. The latter can be used as a shooting parameter in order to verify the last boundary condition at infinity. 

In fact this last boundary condition can not be implemented numerically easily. Therefore, the system \ref{eq:lub}-\ref{eq:CAL} is transposed to curvilinear coordinates $(s,\theta (s))$ and equation \ref{eq:inner} is used to compute the boundary conditions of the system close to the contact line at $z=z_0<<\lambda$. The system then becomes:

\begin{equation} \label{eq:lubcurv}
\left\{
\begin{array}{c}
  \dfrac{\theta''}{\cos \theta}+1 =3\Ca_p\dfrac{1-\dfrac{z}{h}\phi}{h^2+3\lambda h} \\
\partial_s z =\cos \theta \\
	\partial_s h = \sin \theta
\end{array}
\right.
\quad\text{with}\quad
\left\{
\begin{array}{c}
\theta'=h_\lambda''(z_0)\\
\theta=h_\lambda'(z_0)\\
z(s=0)=z_0\\
h(s=0)=h_\lambda(z_0).\\
\end{array}
\right.
\end{equation}

This system is numerically integrated using a custom adaptive stepsize Runge-Kutta fifth order method \citep{Press2007} from $s=0$ up to $s_{\rm{max}}=10$ (to be compared with the unit capillary length) and a dichotomy is used in order to find the value of $C$ such that $\theta(s_{\rm{max}})=\pi/2$. For that value of $C$, it is possible to measure the predicted contact line height $z_{cl}$. Note that the high sensibility of the outcome on the value of $C$ imposes to use a high numerical precision going to 90 digits for decimal numbers. In what follows, the physical slip length is kept constant at an arbitrary value of $\Lambda=1\:\rm{nm}$.

\begin{figure}
  \centerline{\includegraphics[width=0.45\textwidth]{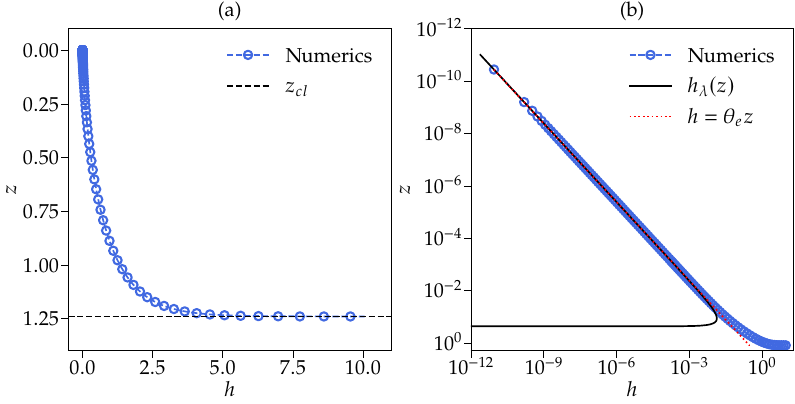}}
  \caption{(a) Computed meniscus profile from Equation \ref{eq:lubcurv} in linear coordinates and (b) in logarithmic coordinates.}
\label{fig:allure}
\end{figure}

On Figure \ref{fig:allure}, the outcome of such a numerical resolution is shown for $\Ca_s=7.4 \times 10^{-7}$, $\Ca_p=2.4\times 10^{-5}$ and $\theta_e=14^\circ$ in linear (a) and logarithmic (b) scales. $z_{cl}$ is indicated on the linear scale graph. On the logarithmic graph, the line of slope $\theta_e$ is also plotted, as well as the asymptotic expression \ref{eq:inner}. We observe that the solutions follows this asymptotic expression, very close here from the line slope up to $h \sim 10^{-3}$ where it departs to join the statuc meniscus profile. 

\subsection{Numerical results}

\begin{figure}
\centerline{ \includegraphics[width=0.45\textwidth]{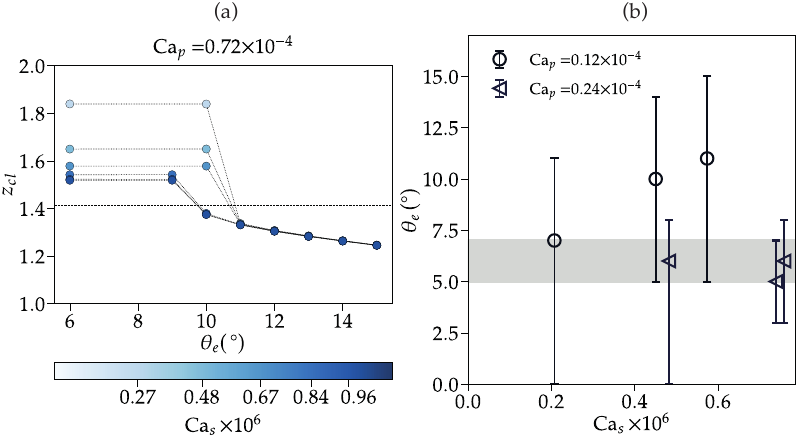}}
  \caption{(a) Predicted contact line height as a function of the input microscopic contact angle for various solidification capillary numbers. (b) Best fit value of the contact angle for all experiments in the sub-$\sqrt{2}$ regime. The grey range between $5^\circ$ and $7^\circ$ represents all possible values of the contact angle between water and ice that would allow to reproduce our experiments within error-bars. }
\label{fig:postheta}
\end{figure}

For a given set of parameters $(\Ca_s,\Ca_p)$, it is thus possible to vary $\theta_e$ in the model \ref{eq:lubcurv} and observe its influence on the value of the predicted $z_{cl}$, which is done on Figure \ref{fig:postheta} (a) for $\Ca_p=7.2\times 10^{-5}$ and different experimental values for $\Ca_s$. Two different regimes are observed: at low values of $\theta_e$, the height of the contact line is $z_{cl}>\sqrt{2}$ and this position seems independent of the value given to the contact angle. Notice however that in these cases $z_{cl}$ still depends on the value of $\Ca_s$, as predicted by equation \ref{eq:scaling2}. For higher values of $\theta_e$, the contact line stays below $\sqrt 2$, and its position strongly depends on $\theta_e$ and does not significantly vary with $\Ca_s$. 

This second regime ($z_{cl}<\sqrt 2$) is of particular interest as it is now possible to compare the experimental measurements in this regime with the numerical prediction in order to obtain a value of $\theta_e$ that can reproduce the experiments. Figure \ref{fig:postheta} shows the best fitting value of $\theta_e$ for all six experiments falling in that regime. Error bars are computed using the error bars on the experimental measurement of $z_{cl}$. Although the data is scattered, all contact angles between $5^\circ$ and $7^\circ$ allow to reproduce our experimental measurement within error bars, suggesting that the effective contact angle of water on ice, when a contact line is receding, should be taken as $\theta_e=6\pm 1^\circ$.

\subsection{Discussion}

\subsubsection{Influence of the geometry}

In the model \ref{eq:lubcurv}, it was assumed that the ice is perfectly planar and grows homogeneously perpendicular to the Peltier element. This is in fact an approximation which neglects the effects of the contact line on the thermal problem. \citet{Anderson1994} indeed proposed that close to a solidifying contact line, the angle between the solid-air and the solid-liquid interfaces should be 90$^\circ$ due to the thermally insulating properties of air. In fact, the analysis of these authors is only valid on a distance $R$ much smaller than the characteristic length $R_0=\sin (U_s/U_p) \frac{D_i}{U_p}\approx \frac{D_iU_s}{U_p^2} \approx 10^{-5}\:\mathrm{m}\approx 10^{-2}l_\gamma$. This would result in curving the ice surface over which the liquid flows, and hence add an additional non dimensionnal curvature gradient of order $1/r_0^2\approx 10^{-4}$. Nonetheless, in the viscous capillary region, the film thickness is expected to be $h<<\Ca_p^{2/3}$. Thus, from equation \ref{eq:lub}: $\kappa'>>\Ca_p^{-1/3}$. Subsequently the correction due to the curved solid-liquid interface is expected to be negligible.

\subsubsection{On the value of the contact angle between water and ice}

The present experiments allow to measure indirectly the contact angle between water and ice in a novel way, as the situation here is strongly out of equilibrium. This approach contrasts with previous attempts seeking for thermal equilibrium \citep{Knight1967, Knight1971, Demmenie2023} or trying to avoid phase change during the measurement \citep{Ketcham1969, Makkonen1997}. Nonetheless, the present approach confirms the shared idea that water does not perfectly wet ice. The physical origin of that unexpected behaviour is less clear. One could argue for instance, in a simplistic manner, that this is due to the anomalous density difference between solid and liquid phase for water, but a complete understanding would need a precise investigation of the surface forces between water and ice, which is out of the scope of the present paper~\cite{Israelachvili}.
Finally, it is worth mentioning that the contact angle measured here is \textit{a priori} not the equilibrium contact angle between water and ice, but rather an effective boundary condition that should be used for the modelling of solidifying receding contact lines, which strongly differs with the previous measurement. This contact angle could be used in order to model more complex unstationary situations such as the spontaneous retraction of thin water films on growing ice layers \citep{Knight1967, Thievenaz2020b}.

\section{Conclusion}
\label{sec:conc}
Thanks to an experimental set-up imposing a constant growth of ice, the withdrawn of a Peltier module at constant speed from a water bath leads to a stationnary regimes where a liquid meniscus is always present, the entrained water being eventually frozen as it is pulled. In this configuration the height of the meniscus contains all the needed information to compute the contact angle of the water on ice in this out of equilibrium regime. By integrating numerically the dynamical meniscus equations in the presence of solidification, we determine that this contact angle is about $6^\circ$ with almost no dependance on the heat flux imposed by the module, suggesting that the contact angle  of water on ice is small but non zero.

\begin{acknowledgments} This work was partially supported by Agence de l'Innovation de D\'efense (AID) - via Centre Interdisciplinaire d'Etudes pour la D\'efense et la S\'ecurit\'e (CIEDS) - (project 2021 - ICING).
\end{acknowledgments}

\bibliographystyle{apsrev4-2}
\bibliography{library}

\end{document}